# Closed-loop coupling of personalised and foundation models for real-time treatment guidance with MRI

**James Grover[1], Emily A. Hewson[1], Andrew Phair[1], Michael Ferraro[1], Hilary L. Byrne[1,2], Paul Keall[1], Michael G. Jameson[2,3,4], and David E.J. Waddington[1]**

[1]Image X Institute, Sydney School of Health Sciences, Faculty of Medicine and Health, The University of Sydney, Sydney, Australia

[2]GenesisCare, Sydney, Australia

[3]School of Clinical Medicine, Medicine & Health, University of New South Wales, Sydney, Australia

[4]Centre for Medical Radiation Physics, School of Mathematics and Physics, University of Wollongong, Wollongong, Australia

Correspondence: james.grover@sydney.edu.au

Image-guided therapies, including radiotherapy, biopsy and deep brain stimulation, rely on real-time targeting of anatomical structures. However, in the presence of motion, imaging latencies create a temporal misalignment between observed and true anatomy, compromising treatment accuracy. Artificial intelligence-based frameworks have increasingly been presented to close this latency gap, but leading personalised models can fail due to a lack of stable anatomical grounding. Foundation models can provide grounded behaviour, but they do not adapt to real-time, individual patient dynamics.

Here we introduce a closed-loop coupling framework that synergises patient-specific temporal prediction with continuous segmentation-based anatomical interpretation from a foundation model. A personalised model predicts future anatomy to compensate for system latency, while a streaming foundation model provides anatomical supervision used to continuously update the temporal predictor in real time during treatment.

We validate the framework using a digital phantom and intrafraction magnetic resonance imaging (MRI) from patients undergoing MRI-guided radiotherapy. For a prediction horizon of 400 ms, the proposed method improves anatomical prediction and reduces dosimetric error compared with existing approaches, within clinically relevant latency constraints. These results establish closed-loop coupling as a general strategy for real-time image-guided intervention.

Artificial intelligence for medical imaging has evolved along two different paradigms: personalised and foundational. Personalised, or patient-specific, frameworks leverage how a patient may be imaged at multiple timepoints throughout their treatment. These frameworks are increasingly popular for treatment guidance in radiation therapy for their ability to capture patient-specific motion dynamics[1-4]. These models can be prone to instability as their training data are limited and patient drift may occur. Foundational artificial intelligence frameworks are also increasingly popular for medical imaging tasks[5]. In contrast to personalised frameworks, foundational frameworks are trained on a large, varied corpus of training data spanning imaging modalities and populations. These frameworks are designed to be broadly applicable across imaging modalities, scanners, and patient populations[5, 6]. A limitation of these frameworks is that they are typically static, not adapting to the patient being imaged on the day. This separation between adaptability and stability fundamentally limits reliable AI-assisted intervention in dynamic settings. Online optimisation of motion prediction models during inference has been explored for centroid-based motion prediction in MRI-guided radiotherapy[7]. Here, we substantially extend this idea: rather than optimising for centroid accuracy, we coupled a personalised temporal prediction model to a foundation segmentation model such that the real-time optimisation is not just driven by image quality but the subsequent segmentation accuracy.

accuracy of delivered treatment[19, 21-23]. Motion prediction can be used to reduce the impact of this latency[24].

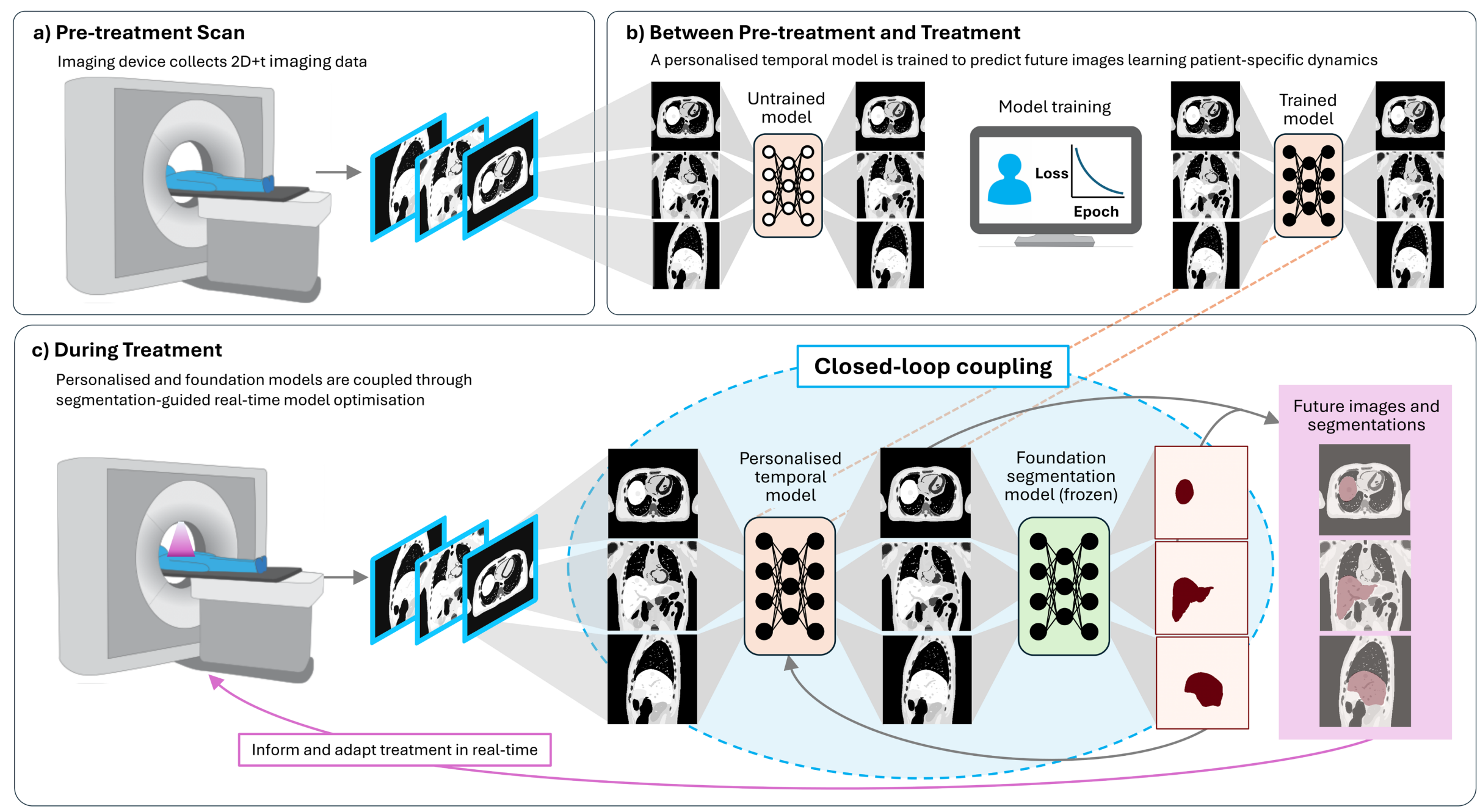

**Figure 1:** Closed-loop coupling of personalised and foundation models for real-time interventions. a) the patient is initially imaged pre-treatment in the treatment position. b) these data are used to train a personalised temporal prediction model to overcome system latencies. c) during treatment, this personalised temporal model produces outputs that are provided directly to a foundation segmentation model. Real-time optimisation of the personalised temporal model based upon the segmentation accuracy of the segmentation model is used to enhance later predictions.

Accurate anatomical targeting is essential for image-guided interventions, including radiotherapy, biopsy, and deep brain stimulation. Real-time MRI is well suited for image-guided interventions due to its superior soft-tissue contrast, enabling continuous visualisation of deformable anatomy during treatment delivery[8]. The adoption of MRI-guided intervention is accelerating, with MRI-linacs now deployed across hundreds of centres worldwide and MRI guidance increasingly used in neurosurgical, cardiac, and biopsy procedures[9-13]. However, respiration induces large, continuous displacements, rotations, and deformations of thoracoabdominal anatomy that must be accounted for in real time to maintain targeting accuracy[14, 15].

Real-time MRI involves trade-offs between the temporal resolution, image quality, and reconstruction accuracy[16, 17]. At clinically used imaging temporal resolutions (~4-5 Hz), MRI-guided radiation therapy systems typically operate with a latency of 300–400 ms, a compromise that balances imaging fidelity against the need for timely anatomical updates[18-20]. Within this latency, anatomy may deviate substantially from observed images, particularly during rapid respiratory motion. This affects both the geometric and dosimetric

Centroid-based motion prediction methods using linear regression and recurrent networks have been deployed clinically but do not capture anatomical rotations or deformations[7, 19, 20, 25-27]. Recently, deep learning has been applied to predict the future full cine MR image or anatomical contour, typically using ConvLSTM-based methods[28-31]. While promising, these methods are restricted to a single imaging plane. This is a fundamental geometric constraint: real-time interventional imaging across MRI, ultrasound and fluoroscopy typically operate on sequential single-plane acquisitions, limiting the amount of anatomical information for guidance. We reason that cross-attention, which learns arbitrary correlations between representations without assuming a fixed geometric relationship, provides a general mechanism for imaging planes to share information and collectively resolve motion that no individual plane can capture alone. We implement this principle in the cross-attention future orthogonal planes (CAFOP) architecture, demonstrated here with orthogonal planes but applicable to any multi-plane acquisition geometry.

Here, we introduce a framework that integrates patient-specific temporal prediction with continuous anatomical supervision from a foundation segmentation model (**Fig. 1**). A personalised CAFOP model predicts future multi-plane MR image triads to compensate for system latency, while the foundation model (MedSAM2[32]) provides real-time segmentation used to update the predictor during treatment. This coupling enables continuous adaptation to intra- and inter-fraction variation while maintaining anatomically consistent predictions.

We validate the proposed framework in the radiation therapy context, where high doses of radiation are delivered to control cancer, using digital phantoms and intrafraction cine MRI from patients treated on a clinical MRI-linac. This validation includes image quality assessment of the predicted images, segmentation accuracy of the tracked anatomical structures, lesion tracking error, and finally, dosimetric accuracy for a beam tracking simulation. The dosimetric results demonstrate that closed-loop AI guidance translates anatomical prediction gains into clinically meaningful improvements in delivered dose, the quantity that ultimately determines treatment efficacy and toxicity. Beyond radiotherapy, the contexts in which latency and patient-specific dynamics jointly degrade targeting accuracy are numerous, spanning MRI-guided focused ultrasound, neurosurgical intervention, cardiac ablation, and real-time biopsy guidance. Closed-loop coupling of personalised and foundation models represents a general strategy for meeting this challenge, and we propose it as a design principle for the next generation of intelligent, adaptive guidance systems in medicine.

## RESULTS

### Cross-attention across imaging planes enables latency-compensated anatomical prediction

We begin by exploring the potential of cross-attention to fuse image information across imaging planes. Our personalised CAFOP architecture was trained on each patient's pre-treatment imaging and evaluated on subsequent images from later treatment fractions, generating latency-compensated predictions for all three orthogonal planes at a 400 ms prediction horizon. Against three established single-plane baselines, ConvLSTM[33], SwinLSTM[34], and SimVP[35], CAFOP significantly outperformed all architectures across image quality and downstream segmentation (MedSAM2[32]) accuracy metrics ($p<0.05$, n=11 patients, Wilcoxon signed-rank test with Holm correction), with the exception of the normalised root mean-square-error (NRMSE) against SwinLSTM (**Fig. 2**). The clinical consequence was most apparent at the liver-diaphragm interface. Pixel sizes were 1.2×1.2 mm² meaning that even small segmentation errors corresponded to millimetre-scale targeting errors that would directly compromise treatment accuracy. CAFOP substantially reduced these errors by anticipating future anatomy rather than acting on delayed observations (**Fig. 2**).

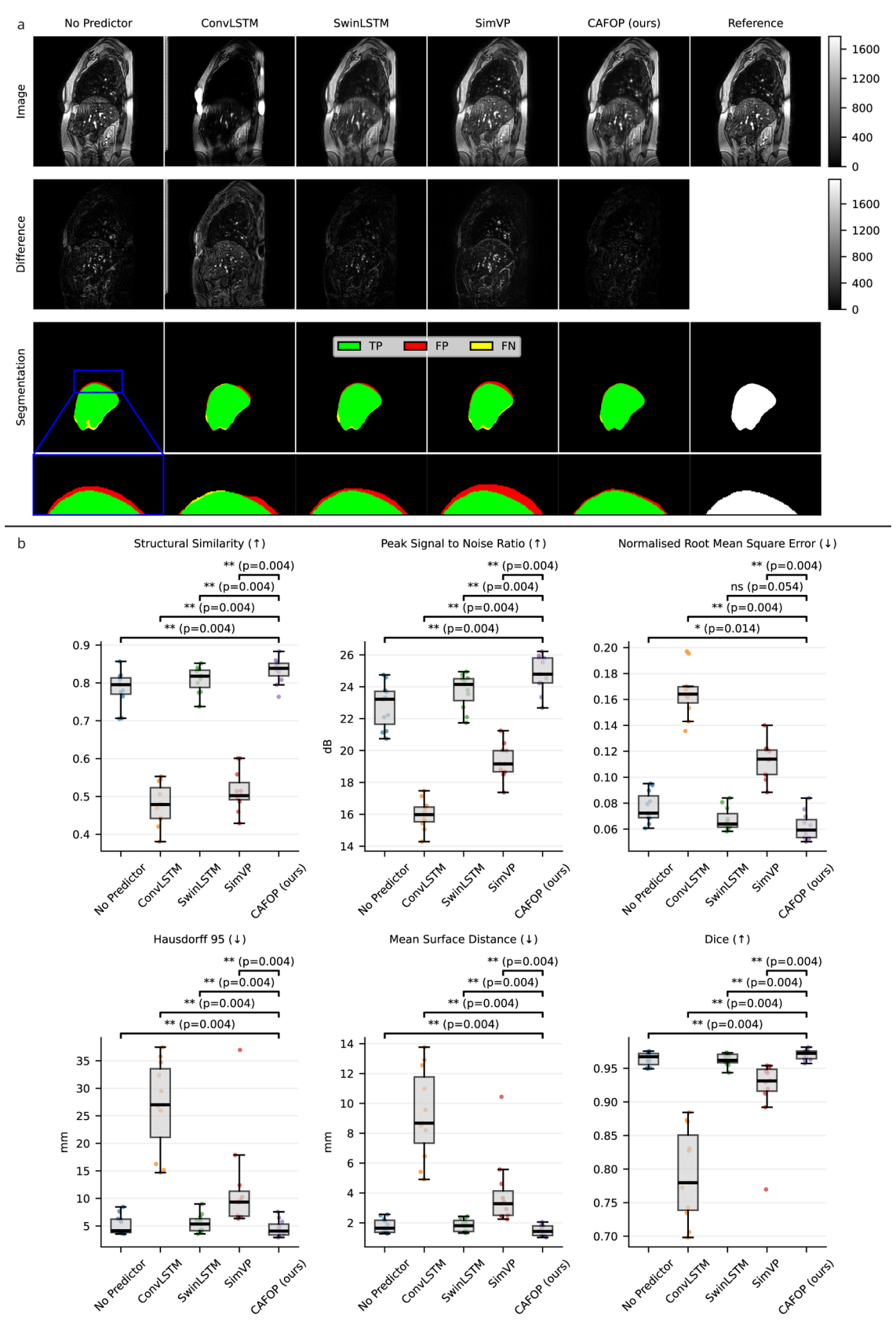

**Figure 2:** Comparison of CAFOP to baseline methods for the task of future image prediction and downstream segmentation. a) example images and segmentations produced by our CAFOP architecture and temporal prediction baselines for the sagittal imaging plane. Pixel sizes were 1.2 × 1.2 mm². Difference maps are all normalised to the same range. TP: true positive; FP: false positive; FN: false negative. b) image quality and segmentation accuracy results between methods. The top row corresponds to image quality metrics and the bottom row to segmentation accuracy metrics. Each data point corresponds to the mean value for a patient. Statistical significance testing was performed using a paired Wilcoxon signed-rank test with Holm correction (n=11 patients). We used the following nomenclature for the star values; p-value ≥ 0.05 as 'ns', 0.01 ≤ p-value < 0.05 as *, 0.001 ≤ p-value < 0.01 as **, and p-value < 0.001 as ***. (↑) corresponds to where a higher metric indicates better performance and (↓) corresponds to where a lower metric indicates better performance.

Seeking to further reduce these segmentation errors, we investigated the concept of closed-loop coupling, whereby the personalised temporal prediction model is directly optimised through the downstream accuracy of the foundation segmentation model.

### Closed-loop coupling improves image quality and segmentation accuracy

Closed-loop coupling continuously re-optimised the personalised temporal prediction model using delayed segmentation labels generated by the foundation model (**Fig. 1**). In this temporal prediction task, delayed segmentation labels were generated by the foundation model and were then used to re-optimise the temporal prediction model for immediate use. This closed-loop coupling significantly improved all image quality and segmentation accuracy metrics for CAFOP ($p<0.05$, n=11 patients, Wilcoxon signed-rank test with Holm correction), even when operating on a delayed stream of data to accommodate the delay in labels being available for optimisation (**Fig. 3**).

While image quality metrics improved by several percent, the largest gains were seen in the Hausdorff 95 and mean surface distance where they improved by approximately 15% using closed-loop coupling (**Fig. 3**). These results demonstrate the benefit of real-time optimisation to downstream tasks such as segmentation and, as we show later, real-time tracking. Specifically, the enhanced personalised temporal model predictions enabled the foundation model to produce more accurate segmentations via this closed-loop coupling mechanism.

Closed-loop coupling is a generalisable concept whereby a personalised model is optimised to a foundation downstream model. This concept is also applied to the other temporal predictions baselines (**Supp. Fig. 1**).

Under this new closed-loop coupling regime, against the baselines, CAFOP outperformed all architectures across image quality and segmentation accuracy metrics at a 400 ms prediction horizon; with improvements reaching statistical significance ($p<0.05$, n=11 patients, Wilcoxon signed-rank test with Holm correction) across all metrics with the exception again of NRMSE against SwinLSTM (**Fig. 4**). Compared to their counterparts in **Fig. 2**, closed-loop coupling quantitatively improved the performance for all models, with substantial gains for ConvLSTM that initially had poor performance. Residual errors were concentrated in the cardiac region, consistent with higher-frequency cardiac dynamics outside the respiratory regime the model is trained to capture. Having demonstrated CAFOP with and without closed-loop coupling paradigms and comparing to baselines, we now focus on the orthogonal views provided by CAFOP that serve to increase the anatomical information available for real-time tracking.

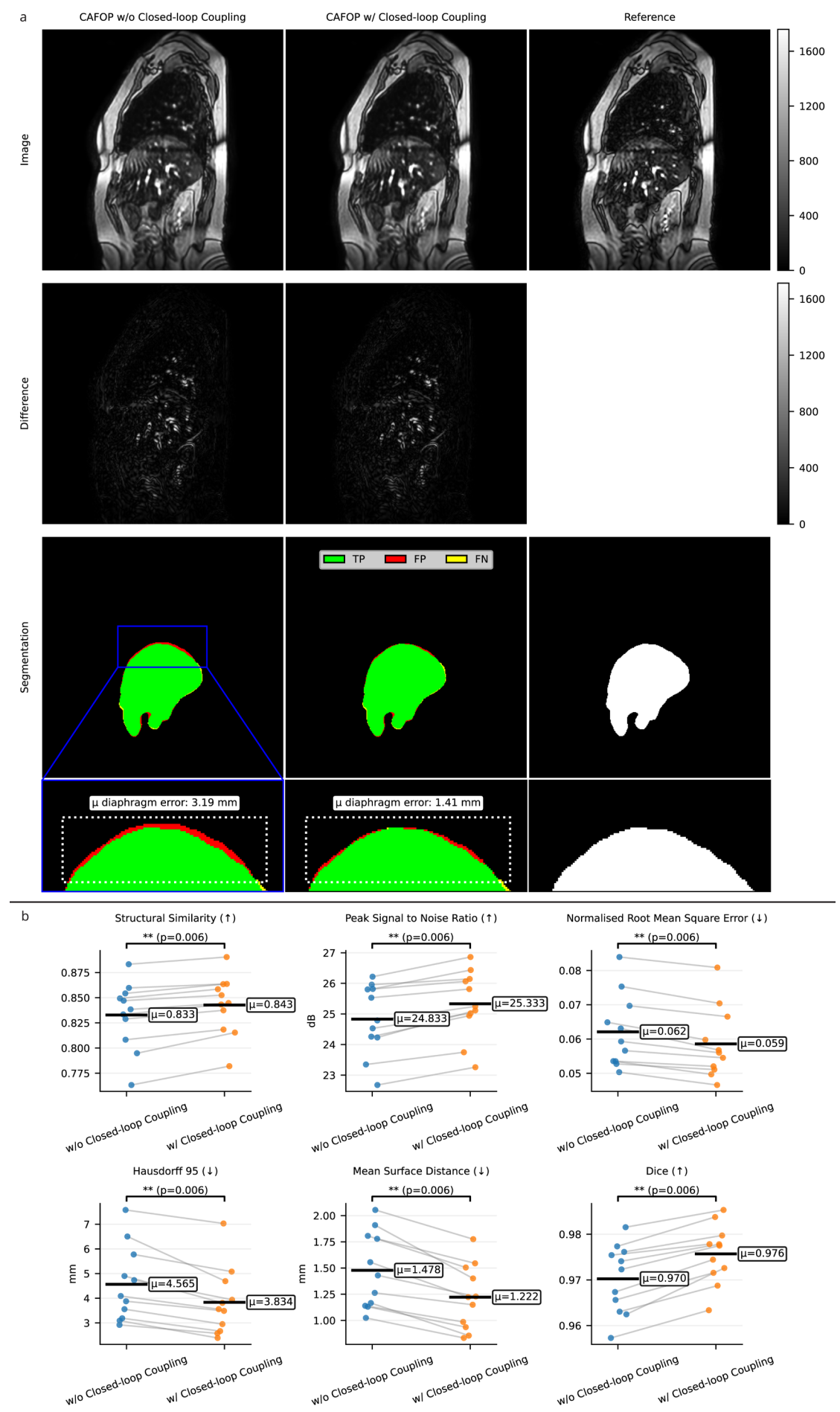

**Figure 3:** Comparison of CAFOP with and without closed-loop coupling. a) demonstrates the differences with and without closed-loop coupling. Pixel sizes were $1.2 \times 1.2$ mm$^2$. Diaphragm error was computed within the bounding box. TP: true positive; FP: false positive; FN: false negative; µ: mean. b) contains the data for all patients, where each data point corresponds to the mean value for a patient. Statistical significance testing was performed using a paired Wilcoxon signed-rank test with Holm correction (n=11 patients). We used the following nomenclature for the star values; p-value $\geq 0.05$ as 'ns', $0.01 \leq$ p-value $< 0.05$ as *, $0.001 \leq$ p-value $< 0.01$ as **, and p-value $< 0.001$ as ***. (↑) corresponds to where a higher metric indicates better performance and (↓) corresponds to where a lower metric indicates better performance. w/o: without; w/: with.

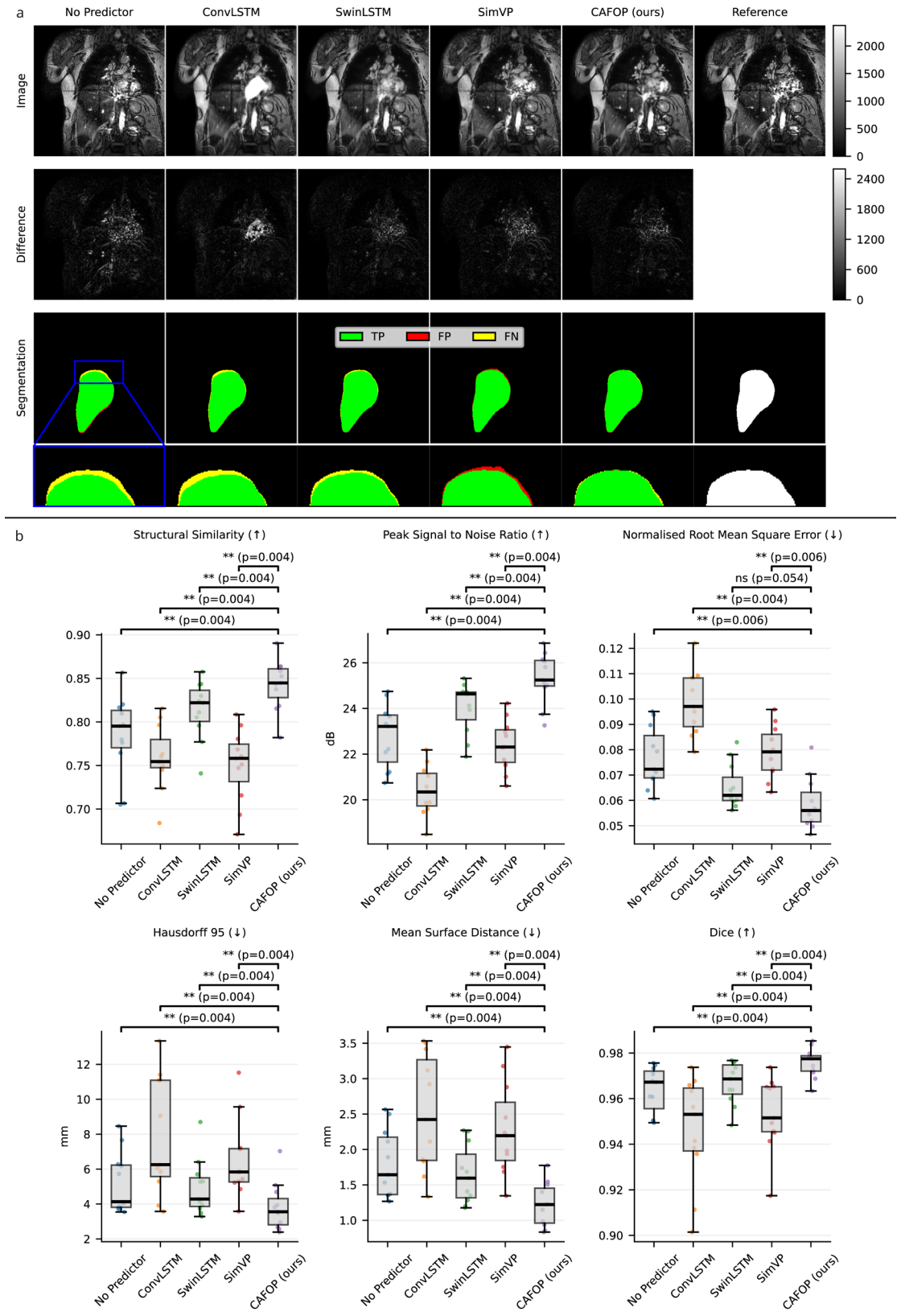

**Figure 4:** Comparison of CAFOP to baseline methods for the task of future image prediction and downstream segmentation using closed-loop coupling. a) example images and segmentations produced by our framework for the coronal imaging plane. Pixel sizes were 1.2 × 1.2 mm$^2$. Difference maps are all normalised to the same range. TP: true positive; FP: false positive; FN: false negative. b) image quality and segmentation accuracy results between methods (all methods used closed-loop coupling). The top row corresponds to image quality metrics and the bottom row to segmentation accuracy metrics. Each data point corresponds to the mean value for a patient. Statistical significance testing was performed using a paired Wilcoxon signed-rank test with Holm correction (n=11 patients). We used the following nomenclature for the star values; p-value ≥ 0.05 as 'ns', 0.01 ≤ p-value < 0.05 as *, 0.001 ≤ p-value < 0.01 as **, and p-value < 0.001 as ***. (↑) corresponds to where a higher metric indicates better performance and (↓) corresponds to where a lower metric indicates better performance.

## CAFOP applies temporal super-resolution for each imaging plane

CAFOP increased the temporal resolution by a factor of three for each imaging plane (**Fig. 5**). Reference imaging on the MRI-Linac is acquired with a temporal resolution of 600 ms per imaging plane due to the interleaved nature of the acquisition. As CAFOP produces images for each imaging plane for each timepoint, this effectively increases this temporal resolution to 200 ms per imaging plane. Timeseries line profiles across each imaging plane demonstrate this temporal super-resolution using closed-loop coupling (**Fig. 5**). Analysis of the diaphragm interface timeseries confirmed that these predictions captured the phase and amplitude of respiratory motion within plausible anatomical bounds.

We additionally observed denoising in the CAFOP-produced images (e.g., the reduction in grainy texture of the liver in the coronal view for **Fig. 5**). We hypothesise that this was due to noise-to-noise effects as CAFOP was trained on noisy data due to limitations in cine-MR acquisition (e.g., no signal averaging due to motion and latency constraints)[36].

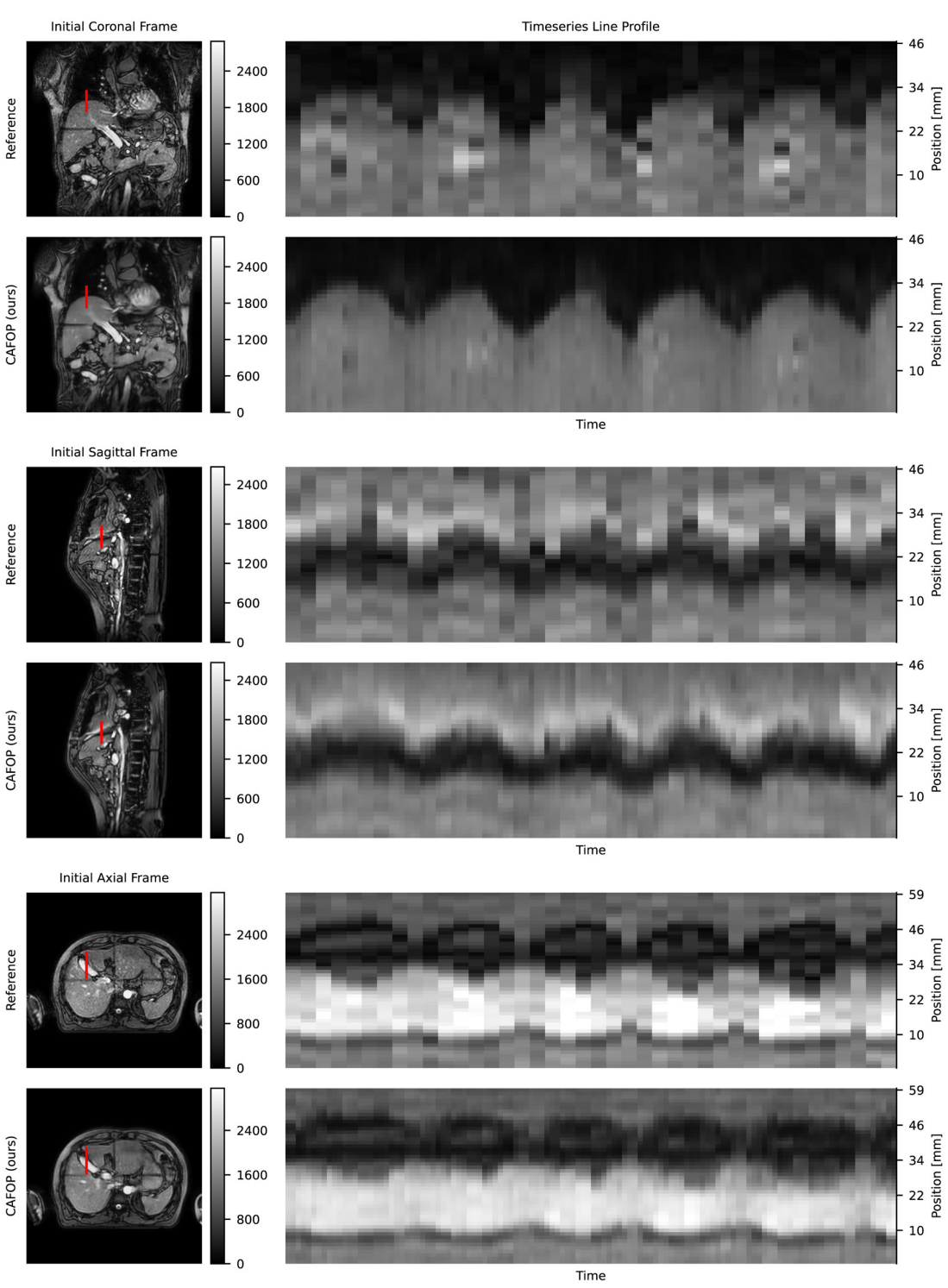

**Figure 5:** CAFOP predictions with reference imaging. The left column shows the initial frames for both the reference and CAFOP prediction. The red line indicates the location of the timeseries profile. The right column shows the timeseries of this line profile for both cine-MR sequences where reference imaging and CAFOP had temporal resolutions of 600 and 200 ms respectively. CAFOP provides temporal super-resolution increasing the granularity of observed respiratory motion while keeping within plausible anatomical bounds as indicated by reference imaging. All images are of the same patient. Pixel sizes were 1.2 × 1.2 mm$^2$.

For every timepoint, a CAFOP prediction contained a single 'on-plane' image and two 'off-plane' images. The on-plane images could be assessed to a reference image while the two off-plane images could not, as the reference acquisition was interleaved. To gain an insight into the quantitative performance of these off-plane images we used a digital phantom with known reference images for all imaging planes (on-plane and off-plane) for every timepoint.

### Digital phantom validates multi-plane prediction and enables real-time tumour tracking

*In vivo* acquisition constrains direct quantitative validation to a single imaging plane per timepoint. A digital phantom (XCAT with CoMBAT modality conversion) with known ground-truth images for all planes and all timepoints was therefore used to assess off-plane prediction quality[37, 38]. Quantitatively, under challenging noise conditions, CAFOP produced images comparable to noiseless reference imaging for all planes (**Fig. 6**). The highest performance was observed for the on-plane predictions, however, off-plane performance was comparable - most notably for the sagittal imaging plane.

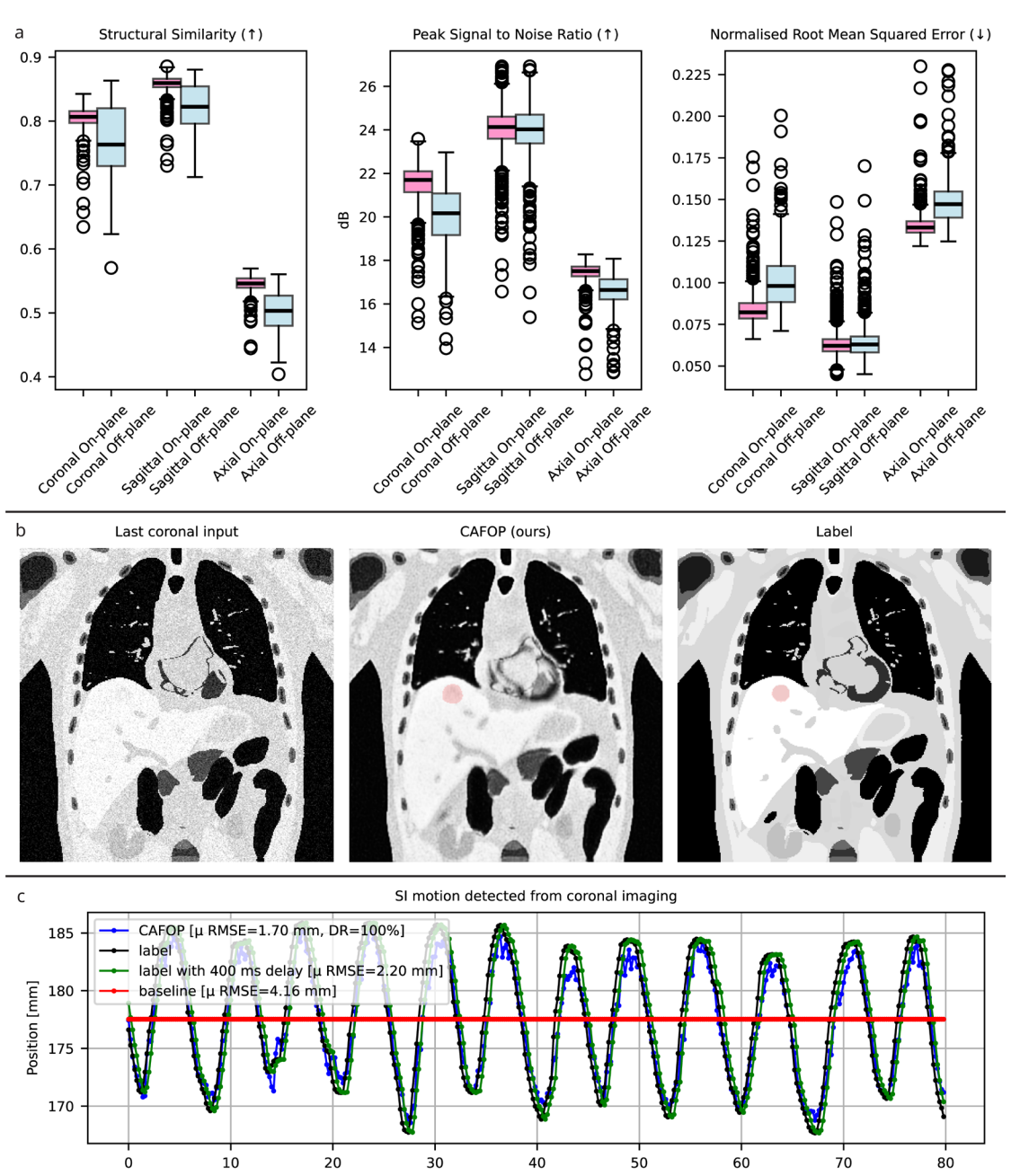

**Figure 6:** Digital phantom (XCAT/CoMBAT) tracking experiment. a) image quality metrics across a simulated fraction stratified by on-plane (referring to where there was a corresponding label for real-time model optimisation) and off-plane (no corresponding label). (↑) corresponds to where a higher metric indicates better performance and (↓) corresponds to where a lower metric indicates better performance. b) example images where the left is the last coronal input to the network demonstrating an obscured tumour due to noise. The centre is the CAFOP prediction for a prediction horizon of 400 ms where the image was denoised and the tumour could be segmented by MedSAM2. The right is the noiseless label image with the tumour also segmented by MedSAM2. c) is a section of the motion trace where tumour was tracked by the label and CAFOP. The label trace was additionally offset by the estimated system latency. The baseline is the mean label superior-inferior position corresponding to a no tracking scenario. The root mean-square-error (relative to the zero-latency label) was computed for all frames in addition to the detection rate (proportion of frames where the tumour could be detected). SI: superior-inferior; μ: mean; RMSE: root mean-square-error; DR: detection rate.

Superior-inferior (SI) motion of a spherical lesion could be tracked with CAFOP and was compared to the noiseless reference tracking (**Fig. 6**). While the baseline sequence had noise levels precluding tracking due to limited lesion/liver contrast, CAFOP with closed-loop coupling was able to successfully detect and localise the lesion. Compared to the baseline of no tracking and in addition to the label centroid position delayed by 400 ms (corresponding to the estimated system latency), CAFOP reduced tracking error by 59% and 23% respectively (down to 1.70 mm). Tracking the lesion in the sagittal plane proved more difficult as demonstrated in **Supp. Fig. 2**, where in some of the frames the lesion could not be localised. While this digital phantom demonstration of real-time tracking was useful in providing insights into CAFOP's performance (with closed-loop coupling); we next performed a beam tracking experiment on clinical treatment data to determine if there was any dosimetric benefit.

### CAFOP with closed-loop coupling achieves consistent dosimetric accuracy

Improved anatomical prediction and segmentation are only clinically meaningful if they translate to more accurate dose delivery. Next-generation radiation therapy is envisioned to have the radiation beam respond to patient motion and the delivered dose in real-time[39]. We used gamma analysis, a standard method for quantifying agreement between planned and delivered dose in radiotherapy, to assess dosimetric accuracy. Typically, failure rates above a few percent in gamma analysis represent clinically meaningful errors for hypofractionated doses. Liver segmentations from each model drove retrospective radiation beam tracking across three patients and six treatment fractions treated at one treatment centre, with liver motion as a surrogate for liver tumour motion. CAFOP with closed-loop coupling reduced mean gamma fail rates from 9.9% with no prediction to 1.5% at the strict 1%/1mm criterion, outperforming all baseline architectures (**Fig. 7**).

Importantly, baseline methods were highly variable across patients with SwinLSTM (the strongest baseline comparator with respect to image quality and segmentation accuracy) exceeding 30% fail rates for individual fractions. CAFOP achieved near-zero fail rates consistently across all patients and fractions analysed at both criteria. The dose difference maps in **Fig. 7** show that CAFOP-enabled tracking produced delivered dose distributions closely matching the planned dose across high-gradient regions, while other methods showed structured geometric offsets translating directly to liver tumour underdosing or normal tissue overdosing. Hence, the image prediction and segmentation gains of the closed-loop framework propagate through to the radiation dose delivered, which is the metric most closely associated with patient outcomes[40].

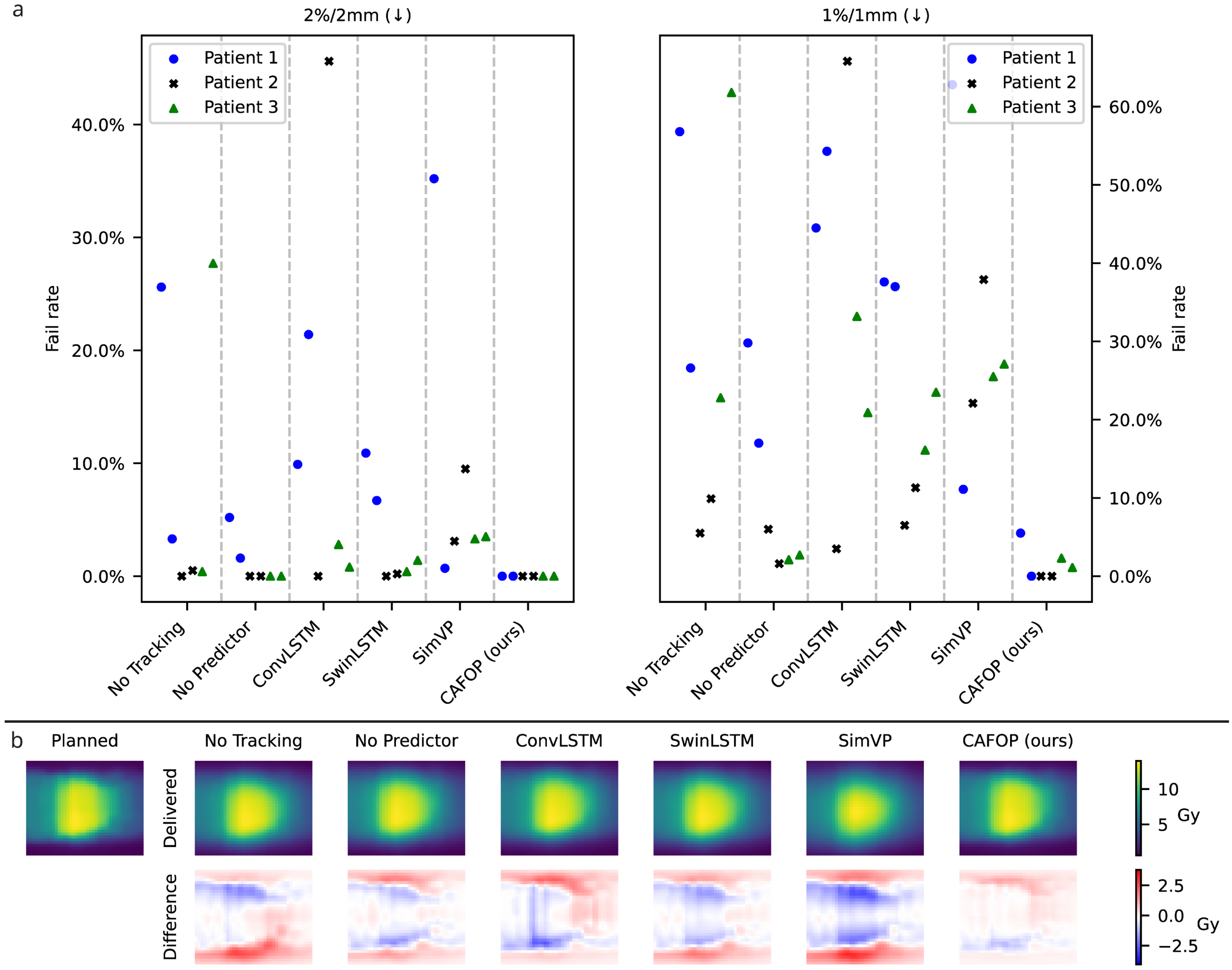

**Figure 7:** Dosimetric results for an in-silico multileaf collimator tracking experiment. a) gamma fail rates between tracking methods. Three patients were analysed each containing two treatment fractions. The left panel corresponds to a 2%/2mm gamma passing criterion and the right panel corresponds to a stricter 1%/1mm criterion. (↓) corresponds to where a lower metric indicates better performance. b) planned, delivered, and difference dose maps for patient 1's first treatment fraction.

### CAFOP operates within clinical latency constraints for real-time deployment

Having demonstrated that closed-loop coupling translates anatomical prediction gains into dosimetric accuracy, we confirm that the framework operates within the expected latency constraints. CAFOP with closed-loop coupling achieved a total modelling time of approximately 130 ms, comprised of 20 ms gradient-tracked CAFOP forward pass (to allow for backpropagation), 30 ms segmentation (when parallelised), and 80 ms backpropagation time (for closed-loop coupling). These times were measured with closed-loop optimisation running on a 1000 ms delayed cache of data, simulating the real-time clinical environment where labels arrive with latency. Adding these modelling times to the estimated imaging and tracking latency (approximately 300 ms), the overall latency was expected to be approximately 400 ms - the same prediction horizon tested in our work - confirming feasibility for real-time deployment (typically ≤ 500 ms)[24]. As the closed-loop coupling operates on a delayed cache of data, when multi-GPU set-ups are available, backpropagation can be decoupled from the real-time inference pipeline, reducing the modelling latency to 50 ms.

## DISCUSSION

We demonstrate that the closed-loop coupling of a personalised temporal prediction model with a foundation segmentation model resolves a fundamental tension in AI-based real-time imaging-guided intervention. Personalised models adapt to individual patient dynamics but lack stable anatomical grounding; foundation models provide grounded, generalisable behaviour but cannot respond to the patient-specific motion encountered during treatment. Closed-loop coupling eliminates this trade-off by using the foundation model's segmentation accuracy to continuously update the personalised predictor in real-time. Applied to MRI-guided radiotherapy, this approach produced consistent dosimetric accuracy across patients and fractions, translating image prediction gains into the clinical endpoint that determines treatment outcome and toxicity. Beyond radiotherapy, the same tension between adaptability and stability arises wherever real-time guidance must respond to individual patient dynamics while maintaining anatomical consistency, spanning MRI-guided focused ultrasound, cardiac ablation, neurosurgical intervention, and real-time biopsy guidance.

Personalised temporal prediction directly addresses the substantial variation in respiratory dynamics between patients and across treatment sessions that makes population-level models poorly suited to real-time interventions including beam adaptation. A personalised model trained on pre-treatment imaging captures individual motion and is continuously refined during treatment as delayed ground-truth labels become available. Foundation segmentation models complement this by providing anatomically consistent outputs without patient-specific labelling, which would be prohibitively expensive for the large quantity of imaging acquired during image-guided procedures (e.g., up to tens of thousands of cine-MR images are acquired during an MRI-guided treatment course). Closed-loop coupling means the foundation model's segmentation accuracy directly shapes the personalised model's training, directly optimising for the downstream task of segmentation rather than image quality in isolation. A further consequence of this training regime was emergent denoising in CAFOP predictions: trained to predict future images from noisy input images, the optimised prediction corresponded to a denoised image, analogous to the noise-to-noise principle[36]. In the cine-MR context, where signal averaging is precluded by motion and latency constraints, this denoising enabled lesion detection that was not possible from the reference acquisition alone.

Online re-optimisation of personalised models during treatment has been demonstrated for centroid-based motion prediction on MRI-linacs, showing that continuously updated models outperform their offline counterparts by adapting to patient-specific respiratory motion in real-time[7]. Patient-specific deep learning has also been established for real-time target localisation in the current imaging frame, achieving accuracy comparable with interobserver variability across a wide range of anatomical sites and tumour types[2]. Rather than predicting current centroid position or localising anatomy in the present frame, CAFOP predicts future anatomy across three orthogonal imaging planes, directly compensating for system latency, with closed-loop coupling driving real-time optimisation through downstream segmentation accuracy. As geometric accuracy alone is insufficient to establish clinical benefit, we extend our analysis to demonstrate dosimetric improvements, which are mechanistically tied to treatment outcomes and toxicity[39, 41]. MedSAM2-derived models have been shown to be the leading approach for tumour tracking in real-time MRI-guided radiotherapy[42], and the closed-loop coupling approach demonstrated here is extensible as foundation models continue to improve.

While tumour position was not tracked directly but rather inferred from liver centroid motion as a surrogate, this is a standard approach in MRI-guided radiotherapy supported by established correlations between liver/diaphragm and tumour displacement during respiration[43]; direct tumour tracking was precluded by the limited contrast between liver lesions and surrounding parenchyma on the cine-MR sequences used here, a challenge previously noted in the literature[44]. Additionally, segmentation accuracy and dosimetric metrics were computed relative to MedSAM2-derived references rather than manual contours, reflecting the impracticality of contouring tens of thousands of cine-MR frames; results should therefore be interpreted in terms of relative model performance. Banding artefacts inherent to the bSSFP sequence used for cine acquisition were present in both training and reference data and, as no model was explicitly trained to remove these artefacts, all methods reproduced them in prediction. As noted, the closed-loop framework is agnostic to the choice of foundation model and future iterations incorporating physician-derived prompts or fine-tuned segmentation models will directly improve absolute performance without requiring changes to the underlying closed-loop coupling mechanism.

Future work will include the extension of the analyses to more patients including across different treatment sites, with a particular emphasis of sites of high intrafraction motion (e.g., the pancreas and lungs). While data were acquired across two treatment centres, validation across field strengths and system configurations remains an important next step toward clinical generalisation. Integration of the proposed framework on a clinical MRI-linac through technologies such as Reconsocket[45] and Gadgetron[46, 47] would enable prospective testing within existing workflows. This prospective testing could involve dose tracking experiments where delivered dose could be compared to the planned dose without relying on simulations and surrogates[48]. Additionally, further exploration of the closed-loop coupling mechanism is warranted in other image-guided interventional regimes.

The results presented here establish closed-loop coupling of personalised and foundation models as a framework for radiation targeting in radiotherapy, with dosimetric evidence demonstrating that image prediction gains translate reliably into clinical accuracy at the endpoint that matters. The cross-attention mechanism's geometry-agnostic design positions CAFOP for validation beyond the orthogonal configuration demonstrated here. Closed-loop coupling of personalised and foundation models represents a practical strategy for real-time image-guided intervention wherever patient-specific dynamics and system latency jointly limit anatomical targeting accuracy.

## METHODS

### Personalised and Foundation Model Framework

We developed a framework that directly integrates a personalised model trained on a patient's pre-treatment data with a foundation segmentation model (**Fig. 1**). The personalised temporal prediction model learned the patient-specific respiratory dynamics to enable latency-compensated imaging by producing images corresponding to the future anatomy of the patient. These predicted future images were provided to the MedSAM2 foundation segmentation model for video segmentation of organ-at-risk structures[32]. These models were synergised through online model optimisation. Here, we exploited the idea that the label images during real-time inference will eventually be available albeit delayed; it was then possible to continuously optimise the personalised temporal model based on the delayed incoming data. A similar idea has been previously explored in the context of centroid prediction for MRI-guided radiation therapy[7]. We significantly extended this idea by not only optimising the personalised temporal model to image quality but also to downstream segmentation accuracy in what we term 'closed-loop coupling'.

### Imaging Data

Data were retrospectively accessed through the ADAPT-MRL registry approved by St. Vincent's Sydney Human Research Ethics Committee (2020/ETHO 1414)[49]. These data were organised in a hierarchical structure by patient then by radiation therapy treatment fraction. Pertinent to the work presented here were the imaging data from each patient, particularly the beam-monitoring intrafraction cine-MRI. Patients were treated under standard-of-care protocols across two treatment centres (three in centre A, eight in centre B) both using 1.5 T *Elekta* Unity MRI-linacs. During treatment, a cine orthogonal interleaved balanced Fast Field Echo (bFFE) sequence is used to monitor intrafraction motion. Temporally, each imaging plane (axial, sagittal, and coronal) is acquired one-by-one in an interleaving pattern. The voxel sizes were between 1.136×1.136 $mm^2$ and 1.300×1.300 $mm^2$ in the in-plane direction (isotropic) with a 5 mm slice thickness. These beam-monitoring cine-MRIs were provided in their raw serialised data format from the Unity MRI-linac. We developed a custom deserialiser to extract the image and header information (voxel size, acquisition time, etc.) from these files. Importantly, these images are the same as those used for real-time motion monitoring on this clinical system. All images were normalised by clipping non-background voxels between their 0.5 and $99.5^{th}$ percentile to reduce the effect of outliers.

For consistency with our framework, we define the patient's first fraction as the 'pre-treatment' imaging session and the second fraction as the 'treatment' imaging session. We repeated this process for subsequent fractions (i.e., the second fraction is now defined as the 'pre-treatment' session and the third fraction as the 'treatment' session). To maximise clinical transferability, we re-trained all models from scratch where there were multiple 'pre-treatment' sessions, removing any potential dependence on a minimum number of fractions before feasibility. Fractions containing less than 100 cine MR images were discarded from analyses due to insufficient imaging time (< 20 seconds) to train personalised models.

## Cross-attention Future Orthogonal Planes (CAFOP) Architecture

We developed a cross-attention future orthogonal planes (CAFOP) architecture that fuses information from multiple imaging planes to produce predictions for all imaging planes for each timepoint. CAFOP, and all other baseline models, were implemented in PyTorch[50]. An architecture diagram is provided in **Fig. 8**.

each sequence prior to cross-attention. For each imaging plane (e.g., coronal), the corresponding sequence is defined as the query ($Q \in \mathbb{R}^{S \times B \times d_{model}}$) and another imaging plane (e.g., sagittal) is defined as the key-value ($K \in \mathbb{R}^{S \times B \times d_{model}}, V \in \mathbb{R}^{S \times B \times d_{model}}$). Q and KV are normalised by separate layer normalisation layers prior to attention computation. The standard multi-head attention layer is applied to these normalised layers producing an attention

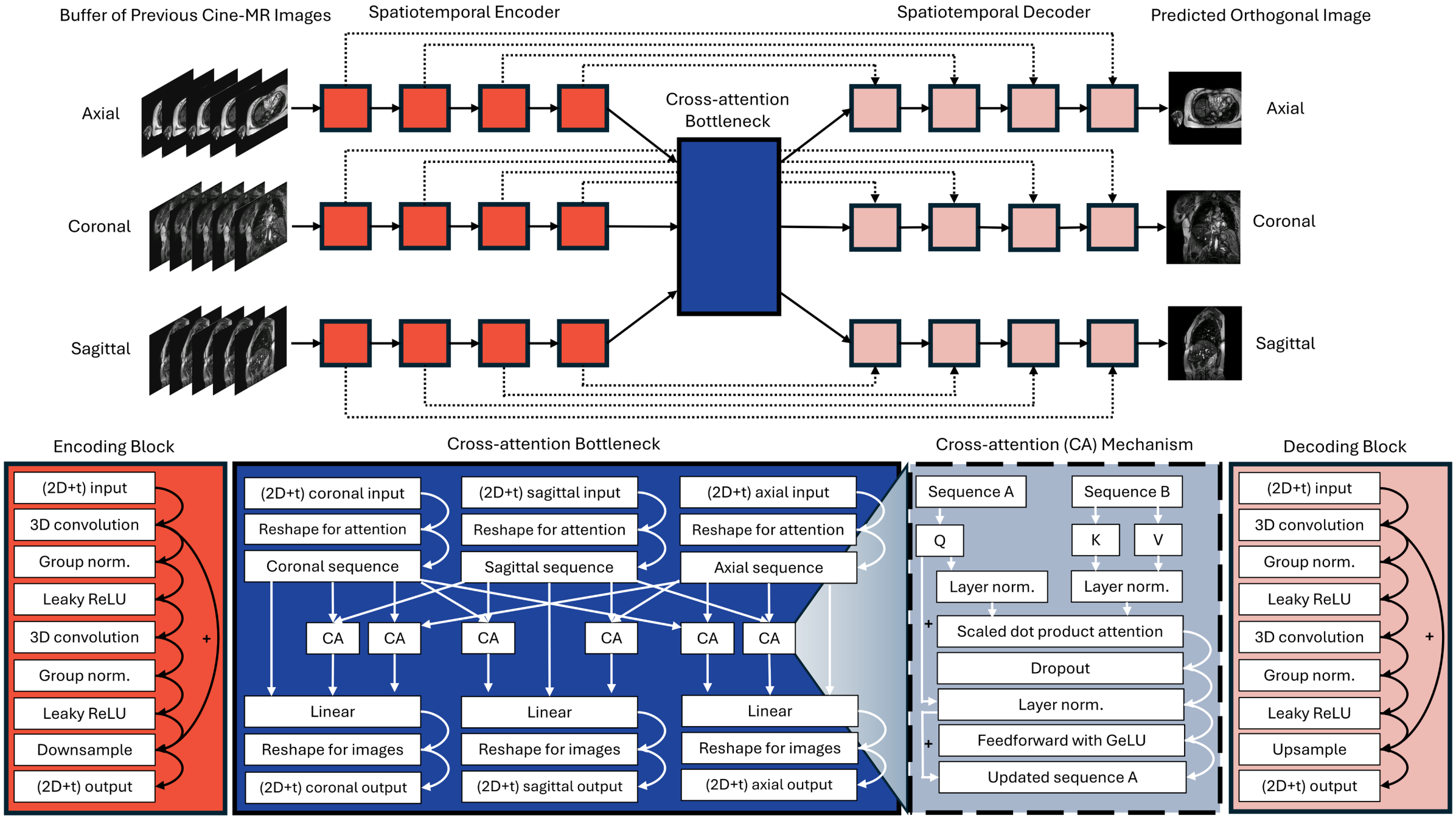

**Figure 8:** Cross-attention future orthogonal planes (CAFOP) architecture. Input buffers of previous cine MR images from each imaging plane are provided to separate convolutional spatiotemporal encoders to learn latent features of each sequence. These encoded images are converted to a sequence and each imaging plane applies cross-attention to each other to directly embed information across planes. These sequences are converted back to images and are decoded by separate spatiotemporal decoders. The final output is an orthogonal image triad corresponding to the future anatomy of the patient.

We explain our architecture with the following notation where $B = \text{batch size}, C = \text{channels}, T = \text{timesteps}, H = \text{height}, W = \text{width}$. The input to our architecture is comprised of a buffer of cine MR images for each imaging plane. The cine MR images for each plane ($I \in \mathbb{R}^{B \times 1 \times T \times H \times W}$) were provided to three independent spatiotemporal encoders where a latent representation was learned for each imaging plane. These spatiotemporal encoders applied 3D convolutions followed by bilinear downsampling (only in the spatial dimensions [H and W]). These latent representations ($\check{I} \in \mathbb{R}^{B \times C \times T \times \frac{H}{16} \times \frac{W}{16}}$) were passed to the cross-attention bottleneck. At the bottleneck, each latent image representation $[B, C, T, \frac{H}{16}, \frac{W}{16}]$ were reshaped to a sequence $[S, B, d_{model}]$ where $S = T \times \frac{H}{16} \times \frac{W}{16}$ and $d_{model} = C$ for the subsequent attention mechanism. Learnable positional encodings were applied to output ($A \in \mathbb{R}^{S \times B \times d_{model}}$). A is then supplied to a feedforward linear network with GeLU activation and dropout for further feature extraction as the original implementation of the transformer encoder[51]. This process repeats for the other imaging plane (e.g., axial as the new key-value). At this point, for each imaging plane there are three corresponding sequences: the original sequence (e.g., coronal) and one for each cross-attention (e.g., coronal-sagittal cross-attention and coronal-axial cross-attention). These are concatenated together ($M \in \mathbb{R}^{S \times B \times 3d_{model}}$) and provided to a linear layer to fuse features extracted from each sequence. This produces a final single sequence for each imaging plane ($N \in \mathbb{R}^{S \times B \times d_{model}}$) that is then reshaped back into the latent image domain ($\check{I} \in \mathbb{R}^{B \times C \times T \times \frac{H}{16} \times \frac{W}{16}}$). Three separate spatiotemporal decoders are used to upsample these latent representations and produce the final predicted images for each imaging plane. These spatiotemporal encoders apply 3D convolutions followed by bilinear upsampling (only in the spatial dimensions [$\mathrm{H}$ and $\mathrm{W}$]). The output of each spatiotemporal encoder block is concatenated to their corresponding spatiotemporal decoder block following the design of U-NET[52]. The resulting output of the model are three future orthogonal images.

### Unsupervised Temporal Interpolation for Orthogonal Views

Given the cine data are acquired using an orthogonal interleaved sequence, there are only data for one imaging plane for a given timepoint. We employed temporal interpolation to generate training data for our CAFOP architecture for it to output three orthogonal imaging planes for a given timepoint. Specifically, we used UVINet – an unsupervised network for temporal interpolation in medical images designed to operate on limited training data[53]. We adapted the UVINet architecture to operate on 2D+t images as opposed to the original implementation operating on 3D+t images. We stratified our orthogonal interleaved sequence into each imaging plane, each with a 600 ms temporal resolution. We trained UVINet to learn a flow between each of the measured images per the original implementation. Once converged, we sampled this learnt flow for each plane every 200 ms to generate training labels for subsequent training of the patient-specific model prior to treatment.

### Baseline Architectures

We implemented three models (ConvLSTM[33], SwinLSTM[34], and SimVP[35]) to compare our novel method to the established methods in the literature. ConvLSTM and SwinLSTM have been directly applied to future cine-MR image prediction[29-31]. SimVP was recently proposed for natural video prediction using a simple convolutional structure as opposed to complex gated mechanisms seen in LSTM-style models. These models were initially developed for natural video prediction and do not incorporate any information sharing between orthogonal imaging planes (c.f., our architecture). Accordingly, we trained and evaluated these baselines on a per-plane approach, where independent models were trained on coronal, sagittal, and axial imaging data. As the temporal resolution for each imaging plane was 600 ms these models were trained for a 600 ms prediction horizon. As the prediction horizon tested in our implementation was 400 ms, corresponding to the estimated latency, a model selection module was developed. This module selected the corresponding imaging plane model based on what the label imaging plane would be for the 400 ms prediction horizon. We use their standard architecture hyperparameters (e.g., number of layers) as described in the literature where they are originally introduced[34, 35], with the exception of ConvLSTM, where we use the implementation described by Lombardo *et al*. in an MRI-guided radiation therapy context[30]. During inference the three models were loaded and individually applied depending on what the next imaging plane should be in the sequence. The loss function (except the temporal change component described later), optimiser, learning rate scheduler, maximum number of epochs, and input time interval (three seconds) were consistent with our proposed architecture.

### Personalised Model Training

The personalised temporal prediction models were trained on the images from the 'pre-treatment' imaging session. As the data were from fractionated radiotherapy patients, we cascaded through their treatment fractions, defining the previous fraction as 'pre-treatment' and the next fraction as 'treatment'. Models were optimised only for reference image quality through a weighted loss function defined in **Eqn. 1**,

$$\mathcal{L}_{\mathrm{Total}} = \lambda_{\mathrm{L1}}\mathcal{L}_{\mathrm{L1}} + \lambda_{\mathrm{SSIM}}\mathcal{L}_{\mathrm{SSIM}}, \tag{1}$$

where $\mathcal{L}_{\mathrm{L1}}$ is the L1 loss and $\mathcal{L}_{\mathrm{SSIM}}$ is the structural similarity loss. $\lambda_{\mathrm{L1}}$ = 0.002 and $\lambda_{\mathrm{SSIM}}$ = 0.998 and were chosen heuristically, optimising global image error and structural fidelity. Models were trained using an adaptive momentum optimiser and a one-cycle learning rate scheduler[54, 55]. The batch size for all models was set to one. Spatial augmentations were applied including random horizontal flipping, rotations, and erasing. The training data were randomly split into 80:20 train:validation split to monitor convergence. Each model was given up to 50 epochs to reach convergence, defined by the mean validation structural similarity not improving for ten consecutive epochs. Once this estimate of the number of epochs was ascertained, the model was trained for the corresponding number of epochs on the full dataset (i.e., no validation). This strategy enabled us to use all the training data while having an indication of when convergence would occur to avoid overfitting. Training time was proportional to the number of cine MRIs and took in the order of hours to days on an NVIDIA A6000 GPU.

## Foundation Model Integration

While images are useful for a variety of purposes for real-time MRI-guided procedures, complementary information can be gained from auto-segmentations. These segmentations can be used for downstream technology such as organ-at-risk monitoring during radiation treatment delivery. MedSAM2 is an attention-based foundation medical image segmentation model that was developed for segmenting medical images in temporal (i.e., 2D+t) datasets[32]. MedSAM2 was chosen due to its generalisability and ability to be fine-tuned to specific anatomical sites or tracking scenarios. Additionally, MedSAM2 operates with a 'mask bank' enabling it to persist segmentations across frames even if intermediate frames have the tracked object fall outside of the field-of-view (a common problem when tracking tumours with 2D+t cine-MR imaging due to out-of-plane motion). MedSAM2 requires an initial mask to be propagated through the temporal sequence. We used MedSAM to generate these initial masks using a rectangle bounding box prompt around the anatomy of interest from the first cine MR frame of the sequence[5]. The base parameters ('medsam_vit_b.pth') were used for all experiments. We generated masks for the liver with a particular focus on the superior dome of the liver at the diaphragm interface as this is tightly coupled with respiratory motion. We note that the initial mask could be generated through other methods such as the radiation oncologist or registration-based methods from planning structures. For real-time model optimisation, we adapted MedSAM2 to operate in a streaming mode, removing the dependence of having the entire sequence at model instantiation. The base parameters ('MedSAM2_latest.pt') were used for all experiments. Three MedSAM2s were instantiated with these trained parameters for each orthogonal imaging plane and ran independently of each other. Their parameters were frozen for all subsequent use, however, gradients were enabled for closed-loop coupling. Our study focuses on organ-at-risk tracking as opposed to tumour tracking as delineation of the tumour on these cine-MRIs can be difficult in many of the frames, a problem previously encountered in the literature[44].

## Closed-loop Coupling

We exploited the principle that during treatment, although they may be delayed, the ground-truth images will eventually become available, allowing for the use of online real-time model optimisation. The model can continue to learn during treatment based on images as they stream in. While similar concepts have been explored for centroid prediction[7], we significantly expand upon this using images and downstream segmentation from a foundation model in a technique we term 'closed-loop coupling'. The outputs of the personalised models, $\hat{y}_{\text{image}}$ are provided to the respective MedSAM2 models to produce segmentation predictions, $\hat{y}_{\text{segmentation}}$. Once the delayed label image, $y_{\text{image}}$, is available, it can also be provided to its respective MedSAM2 model to produce a label segmentation, $y_{\text{segmentation}}$. The personalised temporal prediction models were updated using a loss function as defined by **Eqn. 2**,

$$\mathcal{L}_{\text{Total}} = \lambda_{\text{image}}(\lambda_{\text{L1}}\mathcal{L}_{\text{L1}} + \lambda_{\text{SSIM}}\mathcal{L}_{\text{SSIM}}) + \lambda_{\text{Dice}}\mathcal{L}_{\text{Dice}} + \lambda_{\text{temporal}}\mathcal{L}_{\text{temporal}}, \quad (2)$$

where $\lambda_{\text{image}}$, $\lambda_{\text{Dice}}$, and $\lambda_{\text{temporal}}$ control the contribution of image, segmentation, and change loss and were set to 0.02, 0.02, and 0.01 or 0 (in the case of baseline models as these always had labels) respectively. The image loss used the weighted contribution of L1 ($\mathcal{L}_{\text{L1}}$) and structural similarity ($\mathcal{L}_{\text{SSIM}}$) loss as described in **Eqn. 1**. To guide accurate downstream segmentation and to effectively couple the personalised and foundation models, a Dice ($\mathcal{L}_{\text{Dice}}$) loss was used to update the personalised model parameters. During real-time model optimisation, our CAFOP architecture did not have labels - UVINet relies on the entire sequence being known in advance to learn a flow. To encourage each newly generated image to induce anatomical change, rather than simply duplicate the previous output, we introduce a temporal change loss ($\mathcal{L}_{\text{temporal}}$) that is defined in **Eqn. 3**:

$$\mathcal{L}_{\text{temporal}} = \frac{1}{3} \sum_{\text{i}\in(\text{axial,coronal,sagittal})} \text{ReLU}(\text{cosine_similarity}(\text{prev}_{\text{i}}, \text{curr}_{\text{i}}) - \text{margin}). \quad (3)$$

Here, $i$ indexes each orthogonal imaging plane. The previous prediction for the imaging plane of interest is cached and used to compute the cosine similarity between the current prediction for that plane. The rectified linear (ReLU) function, in conjunction with a margin parameter, are used to only penalise images that have a cosine similarity over a certain threshold. This margin parameter was set heuristically as 0.95 and mean reduction is used across the orthogonal imaging planes.

We model the delay in labels being available for closed-loop coupling by performing the real-time optimisation on a delayed cache. The delay time was set to 1000 ms in our experimentation to serve two purposes: simulating the time it would take for labels to arrive and the time it takes for backpropagation to occur. While this had a slight negative impact on quantitative metrics in initial testing, it is a more realistic demonstration of our framework while providing sufficient time for model backpropagation to occur.

### Image Quality and Segmentation Accuracy Evaluation

Full reference image quality metrics were computed for where there was a corresponding label. Predicted and label images were divided by their corresponding maximum value for normalisation prior to image quality metric calculation. These metrics included the normalised (min-max) root mean-square-error (NRMSE), peak signal-to-noise ratio (PSNR), and structural similarity (SSIM). Additionally, reference segmentation quality metrics were computed for where there was a corresponding label. These metrics included Dice, Hausdorff 95 (HD95), and mean surface distance (MSD). Image quality metrics were computed using the scikit-image Python package[56]. Segmentation accuracy metrics were computed using the Scipy Python package[57].

A multi-stage pipeline was developed to investigate statistical significance between our proposed model and the baseline methods. For each patient (n=11), we computed the arithmetic mean for each metric among their cine-MR images. This was done to avoid skewing results to patients who had a greater number of fractions and/or cine-MR images per fraction. Then for each of these metrics (six in total), we used the Friedman's test to determine if there were any statistically significant ($p < 0.05$) differences between the models[58]. When there were significant differences between models, we then compared our CAFOP to each other baseline with a post-hoc Wilcoxon signed-rank test with Holm correction. Friedman's test and the Wilcoxon signed-rank test were computed using the Scipy[57] Python package while the Holm correction was computed using the statsmodels[59] Python package.

### Digital Phantom Validation

An important aspect of our proposed CAFOP architecture is the ability to generate orthogonal images for a future timepoint. Due to current limitations in MR imaging (i.e., the interleaving pattern of the cine-MRI sequence), only one imaging plane per timepoint could be quantitatively evaluated against a clinical reference image. To further gain an insight into the predictive capabilities of the framework we employed a digital MRI phantom. Specifically, two 3D+t CTs were generated with the 4D XCAT phantom with a temporal resolution of 200 ms (corresponding to the cine-MR temporal resolution)[37]. A 20 mm spherical lesion was placed in the superior portion of the liver. The first 3D+t CT used the liver motion trace of a patient from the LARK clinical trial[60]. The second 3D+t CT's liver motion trace was from the same patient but for a later treatment fraction. The underlying anatomy between the two 3D+t CTs were the same. The first 3D+t CT was used to train a patient-specific model while the second was used as testing.

The CoMBAT phantom workflow was used to convert the 3D+t CT to 3D+t MRI through measured relaxation values and the signal equation for a bSSFP (TrueFISP) sequence[38]. We used a TR/TE of 10/5 ms and additionally a 70-degree flip angle to simulate cine-MR-like contrast. Orthogonal images were extracted from each volume in the 3D+t where the intersection was placed of the initial position of the lesion. This provided us with an orthogonal image triad at every 200 ms from which we could evaluate our framework. We additionally added Gaussian noise to these images to reduce their signal-to-noise ratio to within the range of 20-30. An uncorrupted noiseless reference was also used for generating metric results (image quality and centroid error) but was not used to train CAFOP (i.e., CAFOP trained on noisy labels in both the pre-treatment phase and closed-loop coupling phase). We ran our framework with minimal modification - only the learning rates were altered due to the distribution shift of the digital phantom. Instead of tracking the liver, we tracked the lesion within the liver using MedSAM2 to generate both label and CAFOP segmentations. From this we could compute the lesion centroid error for every timepoint and we measured the root mean-square-error. We restricted our quantitative analysis here to the superior-inferior direction (i.e., coronal and sagittal imaging planes) as the lesion was prone to move out of the axial imaging plane meaning there were no reference or predicted centroids.

### Dosimetric Evaluation

An assessment of how improved real-time segmentations could impact delivered radiation dose was performed by simulating treatment deliveries using a previously described multi-leaf collimator (MLC) tracking framework[61, 62]. Treatment deliveries were simulated for three liver patients and the doses delivered when MLC tracking was performed with motion predicted using CAFOP, the baseline architectures, and no predictor, were compared to the planned doses using a 3D gamma comparison[63, 64]. Intensity Modulated Radiotherapy plans for the Elekta Unity MRI-linac were generated for three of the patients with a dose prescription of three fractions of 16 Gy each. The first fraction was used to train the model for the second fraction and therefore was not assessed. Planning target volume margins were created using a 5 mm expansion on the gross tumour volume. No internal target volume was used. Liver motion was evaluated by calculating the centroid of the liver segmentation in each MRI image predicted by each of the evaluated models. The 3D displacement vector was updated every 200 ms with the two resolvable dimensions available in the current imaging plane. The 3D displacement vector was input into the MLC tracking framework which calculated the MLC leaf positions to adapt to the latest tumour position. The tumour position was assumed to move rigidly with the centroid of the liver segmentation. MLC tracking was simulated with motion predicted using CAFOP and the baseline architectures with a 400 ms prediction horizon, and without prediction. The delivered doses were accumulated using a finite pencil beam model in a volume undergoing the ground-truth motion derived from liver segmentations observed in the reference images[65]. The dosimetric error of each tracking strategy was determined by comparing the delivered doses to the dose produced by the original treatment plan using a 3D gamma comparison with a 2%/2mm and 1%/1mm pass criterion.

### Modelling Time Analysis

All model training and evaluation (including inference times) were performed on the same performance workstation with an Intel Xeon Gold 6248R CPU, 256 GB RAM, and dual NVIDIA RTX A6000s (a single GPU was used for experimentation). We measure all inference times by way of the internal system clock when the machine was unencumbered. A 100 ms delay was added between consecutive predictions to simulate the down time between consecutive cine-MR images. Our CAFOP model needed three MedSAM2 models, one for each orthogonal imaging plane for each timepoint. While that initially indicated a longer inference time compared to single-plane approaches, these MedSAM2 models could be run in parallel as there is no information sharing between these models. As a result, our segmentation time is the time it took for a single MedSAM2 to perform inference.

## ACKNOWLEDGEMENTS

The authors extend their gratitude to the patients whose data were used in this study. The authors thank Tania Twentyman for assistance with data transfer. The authors thank Julia Johnson for the production of Figure 1. The authors thank Dr. Helen Ball, Dr. Thomas Boele, Dr. Nicholas Hindley, and Dr. Zhuang Xiong for providing valuable feedback on the manuscript.

## FUNDING

J.G. is supported by an Australian Government Research Training Program scholarship and an Australian National Health and Medical Research Council Investigator Grant supplementary scholarship. J.G. and D.E.J.W. acknowledge support from Tour de Cure. P.K. and D.E.J.W. received support from the Australian Government National Health and Medical Research Council Investigator Grants 2041792 and 2017140, respectively. D.E.J.W., E.A.H., and P.K. acknowledge funding from a Cancer Council NSW Project Grant (RG 25-01).

# SUPPLEMENTARY INFORMATION

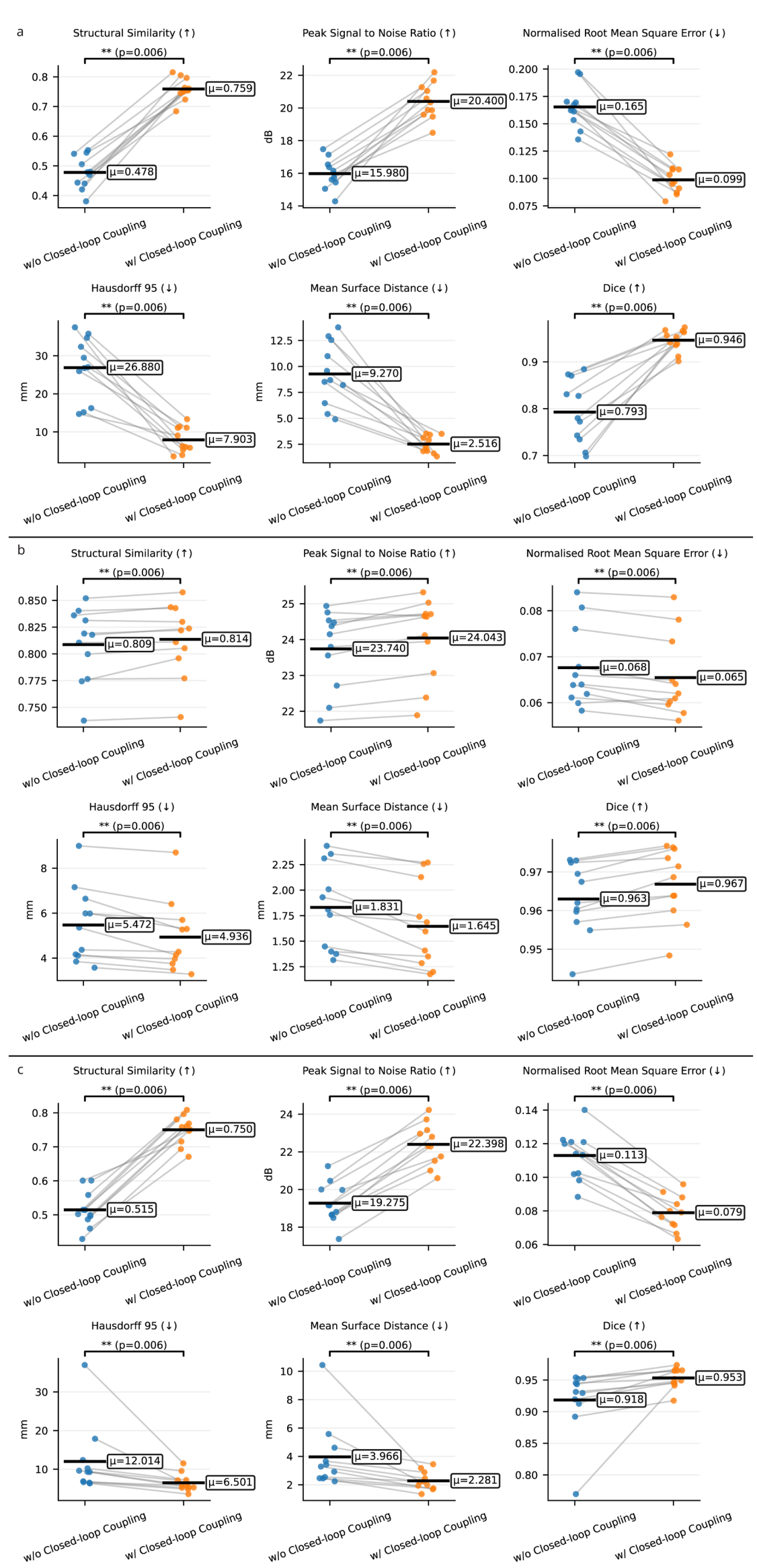

**Supplementary Figure 1:** Comparison of baseline architectures with and without closed-loop coupling. a) ConvLSTM, b) SwinLSTM, and c) SimVP where each panel contains the data for all patients and each data point corresponds to the mean value for a patient. Statistical significance testing was performed using a paired Wilcoxon signed-rank test with Holm correction (n=11 patients). We used the following nomenclature for the star values; p-value ≥ 0.05 as 'ns', 0.01 ≤ p-value < 0.05 as *, 0.001 ≤ p-value < 0.01 as **, and p-value < 0.001 as ***. (↑) corresponds to where a higher metric indicates better performance and (↓) corresponds to where a lower metric indicates better performance. w/o: without; w/: with.

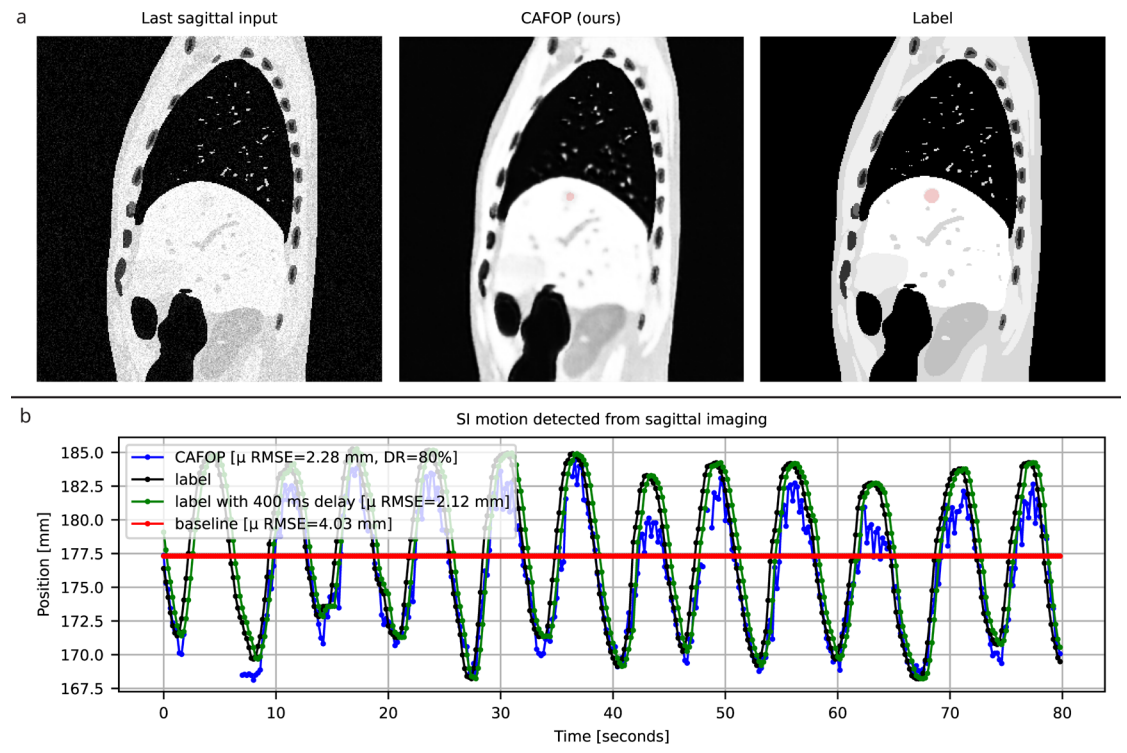

**Supplementary Figure 2:** Digital phantom (XCAT/CoMBAT) tracking experiment. a) example images where the left is the last sagittal input to the network demonstrating an obscured tumour due to noise. The centre is the CAFOP prediction for a prediction horizon of 400 ms where the image was denoised and the tumour could be segmented by MedSAM2. The right is the noiseless label image with the tumour also segmented by MedSAM2. b) is a section of the motion trace where tumour was tracked by the label and CAFOP. There are sections of the motion trace where MedSAM2 failed to localise the tumour. The label trace was additionally offset by the estimated system latency. The baseline is the mean label superior-inferior position corresponding to a no tracking scenario. The root mean-square-error (relative to the zero-latency label) was computed for all frames in addition to the detection rate (proportion of frames where the tumour could be detected). SI: superior-inferior; μ: mean; RMSE: root mean-square-error; DR: detection rate.